\documentclass[11pt]{article}

\newtheorem{theorem}{Theorem}
\newtheorem{lemma}[theorem]{Lemma}
\newtheorem{corollary}[theorem]{Corollary}

\catcode`@=11
\def\@bindentby#1{%
\begingroup\setbox3=\hbox{#1}\par\noindent%
\dimen2=\linewidth\dimen4=\@totalleftmargin%
\copy3\unskip\advance\linewidth-\wd3\advance\@totalleftmargin\wd3%
\parshape=2\dimen4\dimen2\@totalleftmargin\linewidth%
\everypar{\parshape=1\@totalleftmargin\linewidth}%
\ignorespaces}

\def\@eindentby{
	\advance\linewidth\wd3
	\advance\@totalleftmargin-\wd3
	\everypar{\parshape=1\@totalleftmargin\linewidth}
	\par\endgroup}

\newenvironment{indentby}[1]{\@bindentby{#1}}{\@eindentby}
\catcode`@=12

\newcommand{\Singlespacing}{\baselineskip 12pt}

\newcommand{\bidon}[1]{}

\newenvironment{algorithm}[1]{\Singlespacing\noindent\ \\
  \mbox{{\bf Algorithm {#1}}} \\
  [-0.1in]}{\vspace{-0.1in}\mbox{{\bf End of the Algorithm}}\smallskip}

\newcommand{\alginput}[1]
{\begin{indentby}
  {\protect\makebox[0.125in]{}{\em Input:\ \ \ }}#1
 \end{indentby}}

\newcommand{\algoutput}[1]
{\begin{indentby}
  {\protect\makebox[0.125in]{}{\em Output: }}#1
 \end{indentby} \vspace{-0.1in}}

\begin{document}

\renewcommand{\thefootnote}{\fnsymbol{footnote}}

\title{Circular Separability of Polygons\footnotemark[1]}

\author{Jean-Daniel Boissonnat\footnotemark[2] \and Jurek Czyzowicz\footnotemark
[3]
\and Olivier Devillers\footnotemark[2] 
\and Mariette Yvinec\footnotemark[2]}

\date{\today}

\maketitle

\begin{abstract}
Two planar sets are circularly separable if there exists a circle
enclosing one of the sets and 
whose open 
interior disk 
does not intersect the other set.
 This paper studies
two problems related to circular separability. A linear-time algorithm
is proposed to decide if two polygons are circularly separable. The
algorithm outputs the smallest
separating circle. The second problem asks for the largest circle 
included in a
 preprocessed, convex polygon, under some point and/or line constraints.
The resulting circle must contain the query points and it must lie in the
halfplanes delimited by the query lines.

\end{abstract}

\footnotetext[1]{ {\em This work has been supported in part by the ESPRIT Basic
Research Actions Nr. 7141 (ALCOM II) and Nr. 6546 (PROMotion), NSERC, FCAR and F
ODAR.}
A first version of this paper was published in SODA 1995 }

\footnotetext[2]{ INRIA, 2004 Route des Lucioles, B.P.109, 06561 Valbonne
cedex, France\\ Phone~: +33 93 65 77 38, E-mails~:\ firstname.name@sophia.inria.
fr}

\footnotetext[3]{D\'epartement d'informatique, Universit\'e du Qu\'ebec \`a Hull
}

\renewcommand{\thefootnote}{\arabic{footnote}}

\thispagestyle{empty}
\setcounter{page}{0}
\newpage

\section{Introduction}

Let $\cal C$ denote a family of orientable
surfaces in the Euclidean space $E^d$.
We say that $P \subset E^d$ and $Q \subset E^d$ are $\cal C$-separable, if 
there exists $\Sigma \in \cal C$, such that every point of $P$ lies on one side
of $\Sigma$ and every point of $Q$ lies on the other side. In the last decade,
diverse aspects of the separability problem attracted  research interest,
with $\cal C$
most often being considered as the families of hyperplanes, spheres
and polyhedra. For $P$ and $Q$ being two finite sets of points, the
hyperplane separability may be solved by linear programming \cite{m-lpltw-84}. 
In the case of $P$ and $Q$ being two convex polyhedra, this problem is
efficiently solved in \cite{dk-ladsc-85}. 

The problem of finding a polygon with minimum number of vertices, separating
two finite sets of points was studied in \cite{ep-mps-88}. In 
\cite{abosy-fmcnp-89} the same problem of minimal polygonal separation was
solved for the case of two nested, convex polygons. Das and Joseph
\cite{dj-cmcnp-90} proves that finding a separating polyhedron,
having minimum number of faces for two nested convex polyhedra is
NP-complete. 

In \cite{ms-sapo-95} and \cite{bg-aoscf-95} the problem of finding a 
separating polyhedron with approximatively minimum number
of faces is tackled. 
In \cite{m-idssp-92},
Mount proposes a $O(n \log n)$ algorithm computing an enveloping triangulation
of simple polygons. After such preprocessing, given arbitrary location of
two polygons, the minimum link polygonal curve separating them may be
computed efficiently.

The interest in circular separability was fueled by applications in pattern
recognition and image processing, \cite{ka-dd-84}
\cite{f-spcrd-86}. Notice that for two finite sets of points, following the 
idea of Lay \cite{l-sss-71}, an instance of a spherical separability problem in
$E^d$ may be transformed into a linear separability problem in $E^{d+1}$, using
a stereographic projection.
Kim and Anderson \cite{ka-dd-84} presented a quadratic
algorithm solving the circular separability problem for two finite sets of
points.
Bhattacharya \cite{b-cspps-88} improves this bound to $O(n \log n)$, 
computing the entire region at which
may be centered all the circles separating the two point sets. O'Rourke,
Kosaraju and Megiddo \cite{okm-ccs-86} proposed optimal algorithms, finding
in $O(n)$ time the smallest separating circle, and in $O(n \log n)$ time 
all largest separating circles for two sets of points. They use the paraboloid
transformation to get an instance of a convex, quadratic minimization
problem in three dimensions. 

In this paper we study two types of problems related to circular separability.
In section 3, we propose a linear time algorithm determining whether two given
simple polygons are separable by a circle. The algorithm simultaneously scans
two structures: (1) the list of edges of one polygon,
and (2) a path in the furthest point Voronoi diagram of the vertices of
the other polygon. The resulting separating circle, which is the smallest
possible, is always centered on this path. In section 4, we address a dynamic
version of another circular separability problem. We preprocess a convex polygon
$P$, so that the largest circle inscribed in $P$, subject to some query
points and/or line constraints may be found efficiently. The resulting circle
must contain the query points, and/or it must lie in the halfplanes delimited
by the query lines. Our interest in the problem was motivated by an application
in motion planning, where convex paths of bounded curvature inside a convex
polygon were to be computed
\cite{bcdry-ctbc-94}.

\section{Preliminaries}

Suppose that we are given a set $S$ of obstacles in the plane, and we are 
looking for circles that do not intersect the interior of any of the obstacles.
The largest such circle, centered at a query point $p$, may be found quickly,
if the Voronoi diagram of $S$ has been precomputed.
When a query point $p$ is localized in a Voronoi cell, the obstacle
closest to $p$ is determined, and the largest circle centered at $p$ may be
easily found.

When the set of obstacles are edges of a convex polygon $P$, its Voronoi
diagram,  also called its {\em skeleton}
partitions of $P$ into convex polygonal cells. As each cell of this partition
is adjacent
to an edge of $P$, the skeleton is a tree. This tree, rooted at the
vertex which is the center of the largest circle inscribed in $P$, will
be called {\em skeleton tree} and denoted $SkT(P)$.
A useful way to represent
$SkT(P)$ is by means of a convex polyhedral surface obtained in the
following way. For each edge $e$ of $P$ consider a plane containing $e$, having
45 degrees angle with the plane of $P$, and such that $P$ lies below this
plane. Take the lower envelope of the arrangement of all planes obtained this
way. It forms a convex polyhedral surface which will be denoted $Skel(P)$.
Obviously, $SkT(P)$ is the projection of the edges of $Skel(P)$ 
onto the plane of $P$.

In the following, a circle is said to be internal to a polygon $P$
if it is included  in the closure of the region which is the interior
of $P$.
There exists a standard mapping $\phi$ from circles lying in the
$xy$-plane to  points of the three-dimensional space.
A circle $\Sigma$ of radius $r$, centered at $(x_0,y_0)$ is mapped
to the point $\phi(\Sigma)=(x_0,y_0,r)$. The points on the vertical line, 
passing through $(x_0,y_0)$, are images of the circles centered at $(x_0,y_0)$.
As each such vertical
line intersects $Skel(P)$ in a single point $(x_0,y_0,z_0)$, 
points below $z_0$ represent circles internal to  $P$, and points above
$z_0$ represent  circles intersecting or enclosing $P$. 
In consequence, the question
of finding the largest internal circle centered at a query point $(x_0,y_0)$ 
is equivalent to vertical ray-shooting from $(x_0,y_0,0)$ to $Skel(P)$.

Take a cone originating at $(x_0,y_0,0)$ with vertical axis and 45 degrees
apex angle. The points on the surface of such cone are images  of
the circles passing through $(x_0,y_0)$. The image of the largest 
 circle internal to $P$ and
passing through $(x_0,y_0)$ is the point with the largest $z$-coordinate 
of the intersection of this cone with $Skel(P)$.

The {\em furthest site Voronoi diagram} for a set $S$ of $m$ given sites
$s_1, s_2,..., s_m$ is a partition of the plane into convex regions
$FSV(s_1), FSV(s_2),..., FSV(s_m)$, such that any point in $FSV(s_i)$ is
farther from $s_i$ than from any other site. 
The region $FSV(s_i)$ is non empty if and only if site $s_i$
is a vertex of the convex hull of set $S$,
 all non empty regions $FSV(s_i)$ are unbounded and
their boundaries form a tree. Each of the vertices of this tree is the center
of a  circle enclosing $S$ passing through vertices of $S$,
which hereafter is called a furthest site Voronoi circle
or an FS-Voronoi circle for short.
Except for the smallest circle enclosing $S$
which may pass through only two  points of $S$,
each FS-Voronoi circle passes through at least three points of $S$.

In this paper, the furthest site Voronoi
diagram will be represented by a forest $FSArcs(S)$ in the following
way. The vertices of $FSArcs(S)$ are in one-to-one correspondence
with the arcs of the FS-Voronoi  circles
extending between two consecutive points of $S$
and  smaller than $\pi$.
The roots of $FSArcs(S)$ are the arcs of the smallest circle enclosing
$S$. 
Let us consider an edge $E$ of the {\em furthest site Voronoi diagram}
which is the common boundary of two cells $FSV(s_i)$ and $FSV(s_j)$.
Edge $E$ is the locus of the centers of circles enclosing $S$
and passing through $s_i$ and $s_j$. The endpoints of $E$
are the center of two FS-Voronoi circles $C_-$ and $C_+$ which are
respectively
the smallest and the largest  circles passing through $s_i$ and $s_j$
and enclosing $S$ (with an exception when $s_is_j$ is the diameter
of the smallest circle enclosing $S$).
 If segment $s_is_j$ is a diameter of $C_-$,
we assume w.l.o.g. that the arc $s_is_j$ of $C_-$ joining
counterclockwisely $s_i$ and $s_j$ is smaller than $\pi$.
If segment $s_is_j$ is not an edge of the convex hull
of $S$, the arc $s_is_j$ of $C_+$ includes at least a  point $s_k$ of $S$
and, in the forest $FSArcs(S)$ the arcs $s_is_k$ and $s_ks_j$
of $C_+$ are the children of the arc $s_is_j$ of $C_-$.
If segment  $s_is_j$ is an edge of the convex hull of $S$,
$C_+$ is the line through $s_i$ and $s_j$ and a terminal node
corresponding to the segment $s_is_j$ is the child 
of the arc $s_is_j$ of $C_-$.
Observe that the arcs of a descending path of $FSArcs(S)$ have
monotonically increasing radii.
Obviously, $FSArcs(S)$ has $O(m)$ complexity.

\begin{figure}[hbt]
\center
\unitlength 1cm
\input{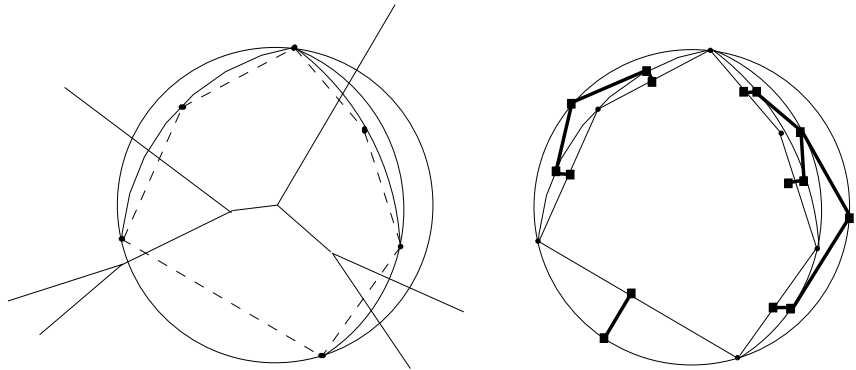}
\caption{The furthest point Voronoi diagram of $S$
and the associated forest $FSArcs(S)$}
\end{figure}

We will use the {\em hierarchical representation} of convex polyhedra
introduced in \cite{dk-ladsc-85}. A hierarchical representation of 
convex polyhedron $D$ is a nested sequence
$D_0 \supset D_1 \supset ... \supset D_k$ of convex polyhedra, such that
(i) $D_0$ is a tetrahedron and $D_k$ is the polyhedron $D$ and
(ii) the set of faces $F_i$ of $D_i$ is obtained from $F_{i+1}$ by removing
a subset $I_{i+1}$ of pairwise non adjacent faces of $D_{i+1}$. 
Polyhedron $D_i$ is then formed from $D_{i+1}$ by extending
remaining faces $F_{i+1} \setminus I_{i+1}$. It may be proved, that
in any polygon $D_{i+1}$ it is always possible to find a set $I_{i+1}$
of $O(|F_{i+1}|)$ faces of bounded degree. Computing of a hierarchical
representation of a convex polyhedron with $n$ vertices may be done
within $O(n \log n)$ time and $O(n)$ 
space. The hierarchical representation supports line
intersection queries in $O(\log n)$ time.

\section{Circles Separating Simple Polygons}

Let $P$ and $Q$ be two simple polygons.
We called the interior of $P$ and $Q$ respectively
the regions bounded
by $P$ and $Q$ denoted $Int(P)$ and $Int(Q)$, respectively.
The regions $Int(P)$ and $Int(Q)$ are considered as open regions.
Let us assume that $P$ and $Q$ have disjoint interiors.
 We say that
circle $\Sigma$ {\em separates} $P$ from $Q$ if  the open disk
which is the interior of $\Sigma$ contains
$Int(P)$ and no point of $Int(Q)$
or vice versa.
In this section, we propose an efficient algorithm to find a circle that
separates two given polygons. The algorithm is designed in such a way, that
it outputs the smallest such circle, or it stops  determining that
no separating circle exists. In some cases, it is possible
that the smallest separating circle has an infinite radius, that is when the
polygons are separable by a line, but not by any finite circle.
The following lemmas specify the condition for  two
polygons to be separable by a circle.

\begin{figure}[hbt]
\center
\unitlength 1.2cm
\input{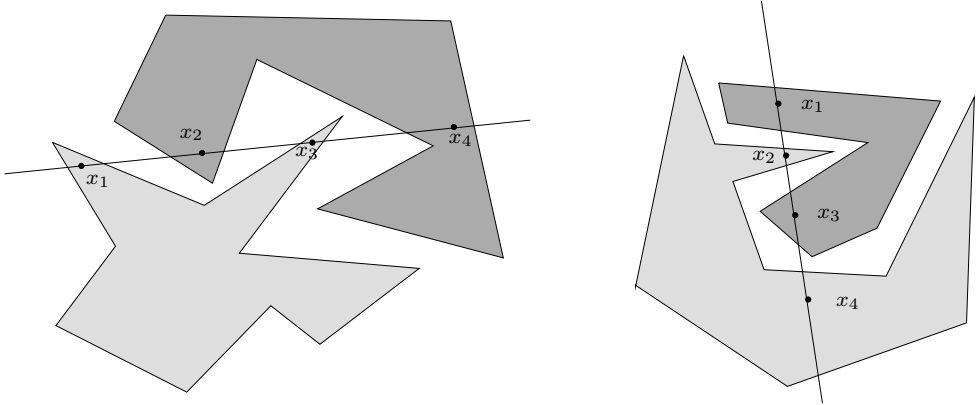}
\caption{There exist four points $x_1$, $x_2$, $x_3$ and
$x_4$, lying in that order on a line, such that $x_1, x_3 \in Int(P)$,
and $x_2, x_4 \in Int(Q)$}
\label{4p}
\end{figure}

\begin{lemma}
\label{4-aligned}
Consider two polygons $P$ and $Q$ with disjoint interiors, such that
$Int(P)\cap CH(Q) \neq \emptyset$ and $Int(Q)\cap CH(P) \neq 
\emptyset$. There exist  a line $l$ and four points $x_1$, $x_2$, $x_3$ and
$x_4$, lying in that order on $l$,
 such that $x_1, x_3 \in Int(P)$,
and $x_2, x_4 \in Int(Q)$ (see Figure~\ref{4p}).
\end{lemma}

{\em Proof~: }
We first define a pocket of $Q$ as a region
of  $CH(Q) \setminus Q$, limited by an edge $E$
of $CH(Q)$ which is not an edge of $Q$ and a part of $Q$
joining the endpoints of $E$.
If $Int(P)\cap CH(Q) \neq \emptyset$, there  exist 
a line $l_1$ and three points $q_1$, $p_3$, $q_2$ in that order
on $l_1$ such that $q_{1},q_{2} \in Int(Q)$ and $p_{3} \in Int(P)$.
Indeed, $Int(P)$ has to intersect at least one of the pockets
${\cal R}$ of $Q$. Then the line going through
a point $p_3 \in Int(P) \cap {\cal R}$ and parallel to the
edge $E={\cal R}\cap CH(Q)$
intersects $int(Q)$ on both sides of $p_3$
and thus is a convenient solution for $l_1$. In the same way,
there is a line $l_2$ and three points $p_1$, $q_3$, $p_2$ in that order
on $l_2$ such that $p_{1},p_{2} \in Int(P)$ and $q_{3} \in Int(Q)$.

\begin{figure}[hbt]
\center
\unitlength 1.2cm
\input{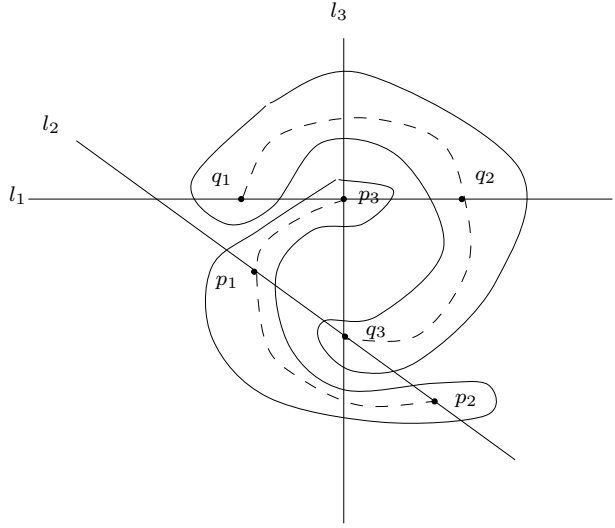}
\caption{For the proof of Lemma~\protect\ref{4-aligned}}
\label{figlemme1}
\end{figure}

Let $l_3$ be the line through $p_3$ and $q_3$ (see Figure~\ref{figlemme1}).
We show now that at least one of the three lines
$l_1, l_2$ or $l_3$ meets the requirement of the lemma.
We note $[q_1,\infty]$ the infinite part  of $l_1$ originating in
 $q_1$ and not including $q_2$. In the same way, we note
$[q_2,\infty]$,
 $[p_1,\infty]$ and 
 $[p_2,\infty]$ the infinite parts of $l_1$ and $l_2$.
Let $i,j \in \{1,2,3 \},\; i \not= j $.
There is a path $\gamma _{q_i,q_j}$
included in $Int(Q)$ and joining $q_{i}$ to $q_{j}$. In the same way,
we shall note $\gamma _{p_i,p_j}$ a path
included in $Int(P)$ and joining $p_{i}$ to $p_{j}$.
Let us assume that  neither $l_1$ nor $l_2$
meets the requirement of the lemma and show that in that case
$l_3$ will do. Since $l_2$ does not meet this requirement,
$\gamma _{q_1,q_2}$ does not intersect  $[p_1,\infty]$
nor $[p_2,\infty]$. Then, we claim that $\gamma _{q_1,q_2}$
 has to intersect $[p_3, \infty]$,
the infinite part of $l_3$ originating in
 $p_3$ and not including $q_3$. Indeed, let $[p_i, \infty]$ with
$ i = 1$ or $2$
 be one of the infinite portions  of $l_2$ that does not intersect
$l_1$.
The concatenation
of $[p_3, \infty]$, $\gamma _{p_3,p_i}$ and  $[p_i, \infty]$
intersects line $l_1$ in the single point $p_3$ and thus
separates $q_1$ from $q_2$. As 
$\gamma _{q_1,q_2}$ cannot intersect $\gamma _{p_3,p_i}$
nor $[p_i, \infty]$
it has to intersect $[p_3, \infty]$.
In the same way $\gamma _{p_1,p_2}$ has to intersect
$[q_3, \infty]$, the other infinite part of $l_3$,
and $l_3$ meets the requirement of the lemma.
\hfill$\diamondsuit$

\begin{figure}[hbt]
\center
\unitlength 1.0cm
\input{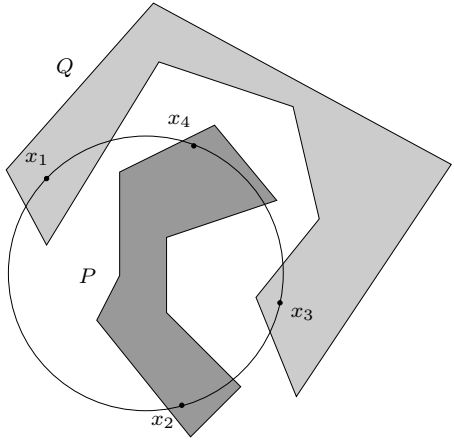}
\caption{No circle separates $P$ and $Q$.}
\label{no-sep}
\end{figure}

\begin{lemma}
\label{exist-circle}
Two polygons $P$ and $Q$ with disjoint interiors cannot be separated by
a circle, if
and only if there exists a circle $C$ and four points $x_1$, $x_2$, $x_3$ and
$x_4$, in that order on the boundary of $C$, such that $x_1, x_3 \in Int(Q)$,
and $x_2, x_4 \in Int(P)$ (see Figure~\ref{no-sep}).
\end{lemma}

{\em Proof:}  We prove first that the existence of a circle $C$
satisfying the
above condition implies that the two
polygons are not separable by a circle. Circle $C$ is split by the points
$x_1$, $x_2$, $x_3$ and $x_4$ into four arcs. Observe that any Jordan curve
$\zeta$ separating $P$ and $Q$ must intersect each of these four arcs.
As any two non-identical circles intersect at two points at most, $\zeta$
cannot be a circle.

\begin{figure}[hbt]
\center
\unitlength 1.1cm
\input{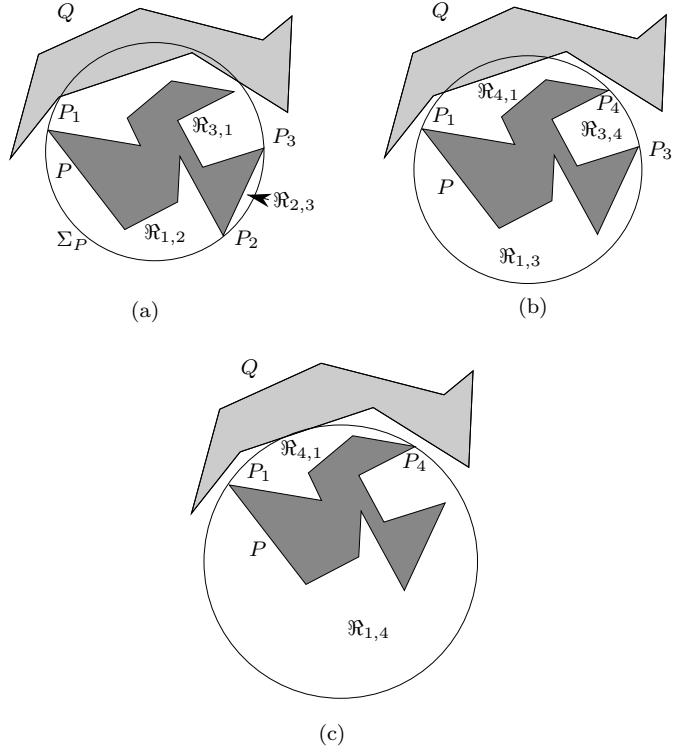}
\caption{Illustrating existence of the separating circle}
\label{sep-poly}
\end{figure}

Assume now that there exists no circle $C$ satisfying the above property.
We prove that $P$ and $Q$ are separable by a circle. Observe first, that
either $Int(Q)\cap CH(P) = \emptyset$ or $Int(P)\cap CH(Q) = \emptyset$,
otherwise, by Lemma \ref{4-aligned}, there would exist four points 
$x_1$, $x_2$, $x_3$ and $x_4$ on a line $l$ 
contradicting our hypothesis. Suppose,
that $Int(Q)\cap CH(P) = \emptyset$, the other case being
symmetrical.
Let $\Sigma$ denote the smallest circle enclosing $P$. 
Consider
$P_1, P_2,..., P_k$, the sequence of points of tangency of ${\Sigma}$
and $P$, in  counterclockwise order around ${\Sigma}$. Denote by
${P}_{P_{i}P_{j}}$ the part of boundary of $P$, extending counterclockwise
from $P_{i}$ to $P_{j}$, and denote by $\Sigma_{P_{i}P_{j}}$ the arc of
$\Sigma$ extending counterclockwise from $P_{i}$ to $P_{j}$. As $\Sigma$
is the smallest circle enclosing $P$, each arc
$\Sigma_{P_{i}P_{i+1}}$ is not greater than $\pi$
(see Figure~\ref{sep-poly}(a)).

Denote by $\Re_{i,i+1}$ the region bounded by ${P}_{P_{i}P_{i+1}}$
and $\Sigma_{P_{i}P_{i+1}}$, $i=1,2,...,n$. The set of regions
$\{\Re_{i,i+1}, i=1,2,...n\}$ constitutes a partition of
$Int(\Sigma) \setminus P$.  If $\Sigma$ does not separate $P$ and $Q$, one
of $\Re_{i,i+1}$ must intersect $Int(Q)$.  Let $P_r$ and $P_s$ denote two 
consecutive points of tangency of $P$ and $\Sigma$, such that $\Re_{r,s}$
intersects $Int(Q)$. Observe that no other region $\Re_{i,i+1}$
intersects $Int(Q)$, otherwise, after shrinking $\Sigma$, we obtain a circle
$C$ having the property mentioned in the lemma.

Continuously increase the radius of circle $\Sigma$, keeping it 
tangent to $P_r$ and $P_s$, until either some new vertex $P_q$ of $P$ becomes
tangent to $\Sigma_{P_{r}P_{s}}$ or until region $\Re_{r,s}$ no longer
meets $Int(Q)$.  In the latter case,
observe
 that at the moment 
$Q$ is  externally tangent to $\Sigma_{P_{r}P_{s}}$ 
(cf. Figure~\ref{sep-poly}(c))
 $Int(Q)$ cannot intersect the opposite region
$\Re_{s,r}$, otherwise the conditions of existence of circle $C$ would be met.
Thus, at that moment the current position 
of $\Sigma$ must separate $P$ and $Q$.
In the former case, the point $P_q$ splits
the arc $\Sigma_{P_{r}P_{s}}$  into two sub-arcs $\Sigma_{P_{r}P_{q}}$
and $\Sigma_{P_{q}P_{s}}$. Region $\Re_{r,s}$ is thus split into two 
subregions $\Re_{r,q}$ and $\Re_{q,s}$. As only one region among $\Re_{r,q}$
and $\Re_{q,s}$, say $\Re_{r,q}$, still intersects $Int(Q)$, 
replace $\Re_{r,s}$ by $\Re_{r,q}$ and continue the process
(cf. Figure~\ref{sep-poly}(b)). Observe that the
radius of $\Sigma$ increases continuously, $\Sigma$ encloses $P$
being tangent in $P_r$ and $P_s$, and arc $\Sigma_{P_{r}P_{s}}$ remains
smaller than $\pi$. As $CH(P) \cap Int(Q) = \emptyset$, 
at some point $\Re_{r,s}$ will no longer intersect $Int(Q)$. 
\hfill$\diamondsuit$

Note that in a special case, when some point of the boundary of $Q$ intersects
the interior of some edge $P_{r}P_s$ of $CH(P)$, the process of increasing
$\Sigma$ stops when the radius of $\Sigma$ reaches infinity. The only
circle separating $P$ and $Q$ will then be a circle of infinite radius, 
being the line of segment $P_{r}P_s$. The following lemma states that, in any
 case, the separating circle found in Lemma \ref{exist-circle}
will be the smallest
possible.

\begin{lemma}
\label{smallest-circle}
If Circle $C$ of radius $r$ intersects polygon $P$ in two points $p_1$ and $p_2$
and polygon $Q$ in point $q$, such that arc $p_1 q p_2$ is smaller
than $\pi$,
any circle enclosing $P$ and separating $P$ from $Q$ must have its radius
greater than $r$.
\end{lemma}

{\em Proof:} obvious.

\subsection{The Algorithm}
 
To determine the separability of two polygons $P$ and $Q$,
the algorithm first looks for the smallest circle
enclosing $P$ and whose interior disk does not intersect $Int(Q)$,
 then looks for the smallest circle
enclosing $Q$  not intersecting $Int(P)$.
For the first purpose, the algorithm 
uses two data structures~: the list ${\cal Q}$ of
edges of polygon $Q$ and the forest  $FSArcs(P)$, 
of arcs of the furthest site Voronoi
circles for the set of vertices of $P$. For any arc
$s_{p}s_q$
of $FSArcs(P)$ and a planar figure $F$ we say that $A$ {\em cuts} 
$F$, if the convex hull of arc $s_{p}s_q$ intersects the interior of $F$. 

The algorithm follows the idea of the proof of Lemma \ref{exist-circle}.
We  first determine an arc $A$ of the smallest circle enclosing $P$
which cuts $Q$. The list $\cal Q$ of edges of $Q$ is then scanned until an
edge $E$ of $Q$ which actually cuts $A$ is found.
A path of a tree of $FSArcs(P)$ is now
traversed until the current arc $A$ admits no children cutting the current
edge $E$. This traversal of $FSArcs(P)$ corresponds to the process
of increasing the radius of the circle enclosing $P$, until 
edge $E$ no longer intersects the circle. 
Then the scanning of list ${\cal Q}$ resumes alternatively with the
traversal of a branch of $FSArcs(P)$ until 
an arc $A$ is found which intersects $Q$ and whose children do not.
Then, let $Arc$ be the arc extending between the endpoints of $A$
and externally 
tangent to $Q$. If 
the circle of $Arc$ does not intersect $Q$, we are done, otherwise
there is no circle separating $P$ and $Q$.

\vbox{
\begin{algorithm}{Smallest Separating Circle}
\alginput{ A simple polygon $P$ of $m$ vertices and a simple polygon $Q$ of $n$ vertices.}
\algoutput{ The smallest circle containing $P$ and disjoint with
	$Int(Q)$, if one exists.}
  \begin{enumerate}
	\item Compute $FSArcs(P)$.
        \item {\bf if} no root of $FSArcs(P)$ cuts $Q$
		\begin{itemize}
		\item[] {\bf then} OUTPUT(the smallest circle enclosing $P$);STOP.
		\item[] {\bf else} $A \leftarrow$ a root of
		$FSArcs(P)$ which cuts $Q$.
		\end{itemize}
  \item {\bf while} ${\cal Q}$ is not empty {\bf do}
        \begin{itemize}
	\item [3.1.] \( E \leftarrow \) next($\cal Q)$
	\item[3.2.] {\bf while }  $A$ does not cut $E$ {\bf do}
                    \begin{itemize}
		        \item[] {\bf if} ${\cal Q}$ is empty, go to 4
			\item[] {\bf else} \( E \leftarrow \) next($\cal Q)$.
                     \end{itemize}
        \item[3.3.] {\bf while }
			there exists a $child_{c}(A)$ which cuts $E$ {\bf do}
	 		   \begin{itemize}
                           \item[] \( A \leftarrow child_{c}(A)\). 
                           \end{itemize}
        \item[3.4]{\bf if} $A$ is a terminal arc of $FSArcs(P)$
		  \begin{itemize}
			\item[] {\bf then} OUTPUT('$CH(P)$ and $Q$ intersect'); STOP.
	           \end{itemize}
	\end{itemize}
  \item $Arc$ $\leftarrow$ the arc externally tangent to
			   $\cal Q$ and passing through the endpoints of $A$.
  \item {\bf if} the complementary arc of $Arc$ cuts the polygon $Q$
	   \begin{itemize}
	      \item[] {\bf then} OUTPUT('$CH(P)$ and $Q$ are not separable').
	      \item[] {\bf else} OUTPUT(circle of $Arc$).
	   \end{itemize}
  \end{enumerate}
\end{algorithm}}

\subsection{The Correctness of the Algorithm}

We prove here that the algorithm outputs the smallest circle
enclosing  $P$ and
external to $Q$ if such a circle exists.

First, we observe that
if the algorithms terminates in step 2, it outputs
the   smallest circle enclosing $P$
which is  clearly the smallest circle enclosing $P$
and external to $Q$ 
if this circle does not intersect $Q$.

Then notice that if algorithm stops with a terminal arc in step 3.4,
the current edge $E$ of $Q$ intersects that terminal arc which is an
edge
of $CH(P)$, thus $CH(P)$ and $Q$ intersect and there is no separating
circle.

If the algorithms does not stop in step 3.4,
the  while loop of step 3 terminates when the list $\cal Q$
is empty. Then,
the current arc $A$
is not a terminal arc and it
cuts $Q$ but its children do not. Indeed, every edge of $Q$
scanned while the current arc is $A$ does not cut  $A$, and
hence these edges do not cut the children of $A$ because
the convex hull
of any arc  contains the convex hull of any 
of its descendant in
$FSArcs(P)$.
Before arc $A$ is the current arc , any scanned edge
was compared with an ancestor of $A$ and found as
not cut by the children of this ancestor of $A$,
therefore the children of $A$ do not cut such an edge.

Let us show that the arc $Arc$ computed in step 4 is uniquely defined.
Let $s_p$ and $s_q$ be the endpoints of the current arc $A$
at the end of step 3.
The segment joining the center of the circle including arc $A$
and the center of the circle including its children
is an edge of the furthest site Voronoi diagram of $P$; this edge is
 the locus of the centers of  circles that
enclose $P$ and pass through $s_p$ and $s_q$.
The arc $s_ps_q$ of the circle including $A$ cuts
$\cal Q$ while the arc $s_ps_q$ 
of the  circle including the children of $A$ does not.
 By continuity, there is a point on this furthest site Voronoi edge  
which is the center
of a circle 
through $s_p$ and $s_q$, enclosing  $P$ and whose arc $s_ps_q$
is  tangent to $\cal Q$. This circle is the extension of $Arc$.

In step 5, when the complementary arc 
$s_{q}s_p$ of $Arc$ cuts the polygon $Q$, there exists a small 
disk $d$ internal to $Q$ and centered in some point $x$ on $s_{p}s_q$. 
Recall that $Arc$ is not greater than $\pi$ and that it is tangent 
at some point $y$ to an edge of $Q$.
Thus it is possible to modify  the circle of $Arc$ slightly, so that it
encloses
point $y$ but neither of $s_p$ and $s_q$, and still
intersects a part of disk $d$. Then,  the condition of Lemma
\ref{exist-circle} is satisfied and there exists no circle
separating $P$ and $Q$. Note that, as step 2 does not compute all the roots
cutting $Q$, and step 4.2 does not test all the children of $A$ for
cutting
 $Q$ the non-separability of $P$ and $Q$ is not detected earlier.

Finally, when the complementary arc of $Arc$ does not cut polygon $Q$,
no edge of $Q$ cuts the interior of the circle
of $Arc$, and the circle encloses $P$. Hence, it is a separating circle.
On the other hand, by the construction of $Arc$, it follows from
Lemma \ref{smallest-circle}, that this circle is the smallest 
separating circle enclosing $P$.

\subsection{The Complexity of the Algorithm}

The first step relies on well known optimal algorithms. By \cite{l-fchsp-83},
the convex hull of $P$ is computed in $O(m)$ time. Within the same complexity,
\cite{agss-ltacv-89} computes the furthest site Voronoi diagram of a
convex polygon, which results in the construction of $FSArcs(P)$
and the smallest circle enclosing $P$.

Step 2 may be computed easily within $O(m+n)$ time in the following way. 
In $O(n)$ time, all edges of $Q$ are tested for intersection with
the interior of the
smallest circle enclosing $P$. Then  any of the edges found to intersect
this disk
is tested for cutting by the roots of $FSArcs(P)$. Since there are $m$ roots
at most, this is done in $O(m)$ time. 

Steps 3.1 and 3.2 are executed at most $O(n)$ times overall,
as each
execution results in skipping an element of $\cal Q$.
 Step 3.3 is executed $O(m)$ times at most, as 
$FSArcs(P)$
 has $O(m)$ complexity. Step 3.4 is executed at most
once. Hence,
the overall complexity of step 3
is $O(m+n)$. 

Step 4 is executed in constant time and Step~5
in $O(m)$ time,
thus
we  conclude with the following result.

\begin{theorem}
\label{2-polygs}
In $O(m+n)$ time and space it is possible to determine whether two given 
polygons, one with $m$ and the other one with $n$ vertices,
are separable by a circle. The smallest separating circle may be
found within the same bounds.
\end{theorem}

Step 5 of the algorithm can be easily extended to exhibit a witness 
(as given by Lemma \ref{exist-circle}) when the two polygons are not separable
by a circle.

Observe that, although it makes no sense to ask for a circle  separating two
polygons with non-disjoint interiors, Algorithm Smallest Separating Circle
still works in this case. The algorithm will either
detect the intersection of the two polygons in step~2
or  stops with a terminal arc in step 3.3. 
The algorithm also works
in the case when polygon $Q$ is not necessarily simple.
Moreover, the algorithm extends to the case when the first
polygonal curve contains the second one, i.e. when we want to separate the
unbounded region lying outside the external curve, from the region bounded
by the internal curve. It is easy to observe that the algorithm
generalizes
also
to the case of separation of connected planar straight line graphs.
 We say that
two graphs are separated by a circle if no edge of the first graph 
intersects the interior of 
the circle while no edge of the second graph intersects the exterior of
the circle. Indeed, in linear time each graph may be transformed to
a polygon, obtained by the traversal of the external face of 
the graph. As some edges may be traversed twice, the polygon is not simple
in general. However,  the algorithm still works in this case.

Furthermore notice that our method can be
extended to answer separability query when
 the allowed separating curves
are the homothets of a given convex curve. Indeed, the algorithm 
relies on Lemma~3 which still holds if the circles are replaced by the
homothets of a given convex curve because two homothets convex curve
intersect in at most two points.
In that case, the algorithm computes
the furthest site Voronoi diagram of polygon $P$ for the convex
distance associated with the given  convex curve.
This can be done in $O(m\log m)$ time, giving a total complexity
of $O(n+mlog m)$.

\section{Largest Circles Inscribed in Convex Polygons}

In this section we study another version of the problem of circular
separability. Suppose that we want to 
separate a convex polygon $P$ from a set of points lying inside the polygon.
Suppose as well, that the polygon $P$ may be preprocessed, so that for each
set $S$ of points given as a query, separation of $P$ from $S$ may be 
decided efficiently. We also address the question when a part of the query 
is the line, delimiting a halfplane in which the separating circle must lie.

\subsection{Point Set Queries}

We start by the case of single point queries.

\begin{theorem}
\label{point}
It is possible to preprocess a convex $n$-gon $P$ in $O(n)$ time and space,
so that given a query point $x$, the largest circle enclosing $x$ and 
internal to $P$ may be found in $O(\log n)$ time.
\end{theorem}

\begin{figure}[hbt]
\center
\unitlength 0.8cm
\input{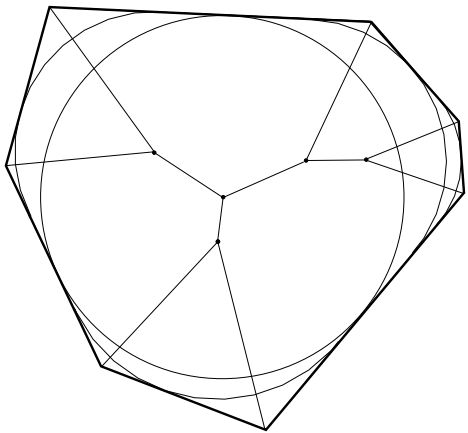}
\caption{Planar map induced by the arcs of Voronoi circles}
\label{map}
\end{figure}

{\em Proof:}
Compute $SkT(P)$ and 
a planar partition of $P$ induced by $SkT(P)$
in the following way. Each vertex of $SkT(P)$ is the center
of a circle internal to $P$ which has at least three tangent points with $P$
and is called a Voronoi circle.
For each Voronoi circle, we consider 
the arcs extending between two consecutive tangent points.
Each such arc which is not greater than $\pi$ is  included in the 
planar map
(see Figure \ref{map}). In this way we obtain a partition
of the interior of $P$.
One region is the interior of the largest circle $C$ inscribed in
$P$. Other regions are bounded by two circular arcs and two parts of 
edges of $P$. Regions adjacent to vertices of $P$ may be considered of
the same type, with one of the arcs degenerated to a single point. As
$SkT(P)$ is computed in $O(n)$ time and space using \cite{agss-ltacv-89}, the 
planar map may be computed within the same bounds. Observe that 
if the query point $x$ lies inside $C$, the largest separating circle is $C$
itself. If point $x$ lies outside $C$, the largest separating circle
passes through $x$ and is tangent to the two portions of edges of $P$,
bounding the region of the map which contains point $x$. Thus, the largest 
separating circle may be found in constant time, once point $x$ has been
located in the planar map. By well-known methods, following the idea of
\cite{k-osps-83}, a trapezoidal decomposition of
our planar map can be preprocessed in $O(n)$ time and
space, so that  point location can be performed in $O(\log n)$ time.
\hfill$\diamondsuit$

\begin{theorem}
\label{large-points}
It is possible to preprocess a convex $n$-gon $P$ in $O(n)$ time and space,
so that given as a query a set $S$ of $k$ points, the largest circle
enclosing $S$ and internal to $P$ may be computed in $O(k \log n)$
time and $O(n+k)$ space.
\end{theorem}

{\em Proof:}
Construct the planar map, as in Figure \ref{map} in the preprocessing step.
$SkT(P)$ is the dual graph of the map.
Let $p$ be a point of $S$. We observe that all maximal disks
included in $Int(P)$ and containing $p$ are centered
on a subtree  of $SkT(P)$ rooted  at the center
of the largest internal circle
passing through $p$. Thus, if two points
$p$ and $q$ of $S$ belong  to two different 
cells of the planar map which
correspond to the unrelated
vertices of $SkT(P)$, i.e. such that neither of these two vertices is an
ancestor of the other one, no circle internal to $P$
contains both points $p$ and $q$. 
Hence, if $S$ is enclosed in a circle internal to $P$,
all points of $S$ must belong to  cells, whose duals belong to a descending
path of $SkT(P)$. To answer the query, we perform first the point location
in the map of each element of $S$. We check next if the cells of the query
points correspond to a descending path in $SkT(P)$. For each query point $q$
we compute the largest circle inscribed in $P$ and containing $q$. The smallest
among all these circles is the candidate for the circle containing $S$.
It is sufficient if all points of $S$ belong to the candidate circle.
The complexity of the algorithm is dominated by the point location step,
taking $O(k \log n)$ time.
\hfill$\diamondsuit$

Remark, that the smallest circle internal to $P$, and containing a set of
$k$ points, may be computed using the technique from the previous section.
The set of $k$ points must first be connected to form the set of vertices of
a polygon. We can conclude by the following alternative result

\begin{corollary}
\label{small-points}
Given a convex $n$-gon $P$ and a set $S$ of $k$ points, the largest circle
containing $S$ and internal to $P$ may be found in
$O(k \log k + n)$ time and $O(n+k)$ space.
\end{corollary}

\subsection{Queries Involving Lines}

We consider first the case when the query consists of a single line,
determining a halfplane which must contain the resulting circle.

\begin{theorem}
\label{line}
It is possible to preprocess a convex $n$-gon $P$ in $O(n \log n)$ time and
space, so that given a query line $l$, the largest circle internal to $P$ and
lying in a closed halfplane $H_{l}^{+}$, determined by $l$,
may be found in $O(\log n)$ time.
\end{theorem}

\begin{figure}[hbt]
\center
\unitlength 0.8cm
\input{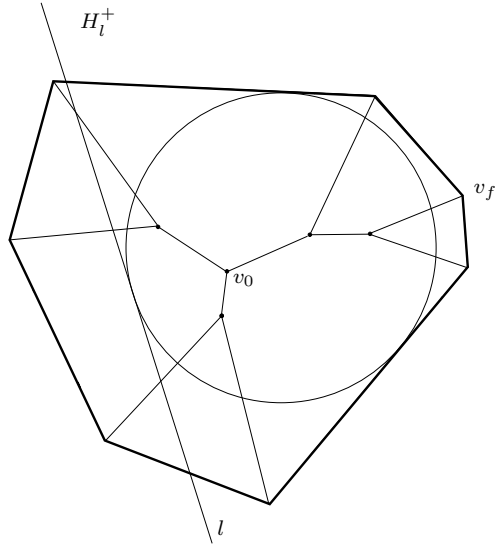}
\caption{The largest circle contained in $P \cap H_{l}^+$ is centered on the
path joining $v_f$ and $v_O$}
\label{ligne}
\end{figure}

{\em Proof:}
Let $v_f \in H_{l}^+$ be the vertex of $P$ which lies at the largest distance
from $l$. The part of the boundary of $P$ lying in $H_{l}^+$ is split by $v_f$
into two chains of edges. The largest circle $C$ inscribed in $P \in H_{l}^+$ 
must be tangent to each of these two chains. $C$ is then centered on the
path of $SkT(P)$ joining its root with vertex $v_f$. See Figure \ref{ligne}.

To answer the query, we first find in $O(\log n)$ time vertex $v_f$. Then we
perform a binary search on the path joining the  root of $SkT(P)$  
 with vertex
$v_f$, to find 
an edge of $SkT(P)$ containing the center of $C$. Now we can find $C$ in
constant time. 

In order to perform above algorithm, an appropriate search structure must be
build in the preprocessing time. It is sufficient to add to each vertex of
$SkT(P)$ the pointers to its ancestors at distance $2^i$, for 
$i=1,2,...,\lfloor \log n \rfloor$. It is possible to construct such
structure in $O(n \log n)$ time and space, during a standard tree-traversal
of $SkT(P)$.
\hfill$\diamondsuit$

Our next result considers the case when the query is given as a pair of
lines, determining a wedge in which the solution circle must be contained.

\begin{theorem}
\label{2lines}
It is possible to preprocess a convex $n$-gon $P$ in $O(n \log n)$ time and
space, so that given as a query two lines $l_1$ and $l_2$, the largest
circle $C$ internal to $P$, and lying in the closed wedge determined by $l_1$
and $l_2$ may be found in $O(\log n)$ time.
\end{theorem}

{\em Proof:}
Three cases are possible. The resulting circle $C$ is tangent to both lines
$l_1$ and $l_2$, it is tangent to one of them, or $C$ does not meet any
of the two lines. Suppose that $C$ is tangent to $l_1$ and $l_2$.
Consider the space of circles introduced in the Preliminaries section.
Take a halfplane ${\cal H}_{l_1}$, originating at line $l_1$ of $x$-$y$ plane,
having 45
degrees angle with the vertical axis. When $C$ is tangent to $l_1$, $\phi(C)$ 
must belong to ${\cal H}_{l_1}$. In our case $\phi(C)$ is the intersection 
of the line $\delta = {\cal H}_{l_1} \cap {\cal H}_{l_2}$ with $Skel(P)$. 
Hence, the problem reduces to finding an intersection of a line with
a convex polyhedron, which may be answered in $O(\log n)$ time,
supposing $O(n \log n)$ computation of the hierarchical representation
of $Skel(P)$ in the preprocessing time.

The algorithm takes four cases into consideration.
In the first case, the largest circle
inscribed in $P$ is output as the solution as long as
it does not intersect $l_1$
nor $l_2$. In the second case, the largest circle contained in 
$P \cap H_{l_1}^+$ is computed. This circle is the solution of our problem
if it does not intersect $l_2$. Similarly, in the third case, 
the largest circle contained in $P \cap H_{l_2}^+$ is computed and then
checked for the intersection with $l_1$. Finally, the largest circle
contained in $P \cap H_{l_1}^+ \cap H_{l_2}^+$ is found using the above
method. Obviously, the solution exists only when 
$P \cap H_{l_1}^+ \cap H_{l_2}^+ \neq \emptyset$.
Except for the first case, our algorithm uses $O(\log n)$ time,
supposing $O(n \log n)$ time preprocessing.
\hfill$\diamondsuit$

A similar
 technique is used to solve the mixed query problem, when the resulting
circle must contain a given point, and it must lie on one side of a given
line.

\begin{theorem}
\label{point_line}
It is possible to preprocess a convex $n$-gon $P$ in $O(n \log n)$ time and
space, so that given  a query consisting of
 a line $l$ and a point $x \in H_{l}^+$,
the largest circle $C$ internal to $P$, enclosing $x$ and
lying in the closed halfplane $H_{l}^+$, may be found in $O(\log n)$ time.
\end{theorem}

{\em Proof:}
Suppose that $C$ is tangent to $l$ and  contains $x$ on its boundary.
$\phi(C)$ lies then on a parabola $\wp$, being the intersection of $H_{l_1}^+$
with the vertical cone originating at $x$, having 45 degrees apex. It is 
possible to adapt the algorithm for line intersection queries to the case
of  the intersections between $Skel(T)$ and  parabola $\wp$. Indeed, 
the parabola  $\wp$ intersects $Skel(T)$
and each polyhedron of the hierarchical decomposition of 
$Skel(T)$ in at most two points. 
To prove the claim, consider the set of circles $\cal C$
passing through $x$ and tangent to $l$. These circles
are centered on the parabola $\wp'$ obtained by projecting $\wp$
onto the $xy$ plane. The claim follows from the fact that
the subset of circles of $\cal C$ that intersect $P$
are centered on a single arc of $\wp'$.

The algorithm checks if the largest circle inscribed in $P$ contains $x$ and
lies in $H_{l}^+$. If this is not the case we find, as in Theorem 
\ref{point}, the largest circle containing $x$, and we output this circle
if it lies in $H_{l}^+$. Otherwise, we continue, as in Theorem \ref{line},
computing the largest circle inscribed in $P$, which lies in $H_{l}^+$.
We output this circle if it contains $x$. Finally, if no circle was
output yet, we find circle $C$ tangent to $l$ and containing $x$ on its
boundary using the above method. The solution does not exist when the
parabola $\wp$ does not intersect $Skel(P)$. The complexity of the query
algorithm is $O(\log n)$. The preprocessing is dominated by the 
$O(n \log n)$ hierarchical decomposition and the construction of
the search structure needed in Theorem \ref{line}.
\hfill$\diamondsuit$

\bidon{
\begin{theorem}
\label{point_line}
It is possible to preprocess a convex $n$-gon $P$ in $O(n \log n)$ time and
space, so that each of the following query can be answered 
in $O(\log n)$ time.
\begin{itemize}
\item Given a query line $l$, find the largest circle internal to $P$ and
lying in a closed halfplane $H_{l}^{+}$ determined by $l$.
\item Given two query lines $l_1$ and $l_2$, find the largest
circle $C$ internal to $P$, and lying in a closed wedge determined by $l_1$
and $l_2$.
\item Given as a query a line $l$ and a point $x \in H_{l}^+$, find
the largest circle $C$ internal to $P$, containing $x$ and
lying in the closed halfplane $H_{l}^+$.
\end{itemize}
\end{theorem}
}

Observe that Theorems~\ref{point}, \ref{large-points},
\ref{small-points}
and \ref{line} can be easily generalized
to queries concerning the homothets of a given convex curve.
In the same way,  Theorems~\ref{2lines} and \ref{point_line}
can be generalized~:  the mapping $\phi$
from the homothet convex curves to points in the three
dimensional space  is defined analogously than in the case
of circles by choosing a reference point internal to the convex curve
and a particular point on the convex curve whose (Euclidean)
distance to the
reference
point will be consider as the radius of the convex curve. Then
the locus of points that are the images of curve internal to $P$
and tangent to $P$ is still a polyhedron $Skel(P)$,
and the locus of points that are images of curves tangent to a line $l$
is still an hyperplane ${\cal H}_l$.
The cone  which is the image
of the convex curves passing through a point $x$ 
is no longer a circular cone but a cone
whose sections perpendicular to the vertical axis
are the homothets of a convex curve
dual to the given  
 convex curve.
The complexity results have to be adapted depending on the complexity 
of the new basic operations used in the algorithms.

\section{Conclusion and Open Problems}

The paper studied two types of problems concerning circular separability. In
Section 3,  the problem of separability of two simple polygons is solved.
Section 4 concerns the problem of the largest circle inscribed in a convex
polygon, given some query 
point and/or line constraints.  The natural way to approach
the circular separability problems is to employ some mixture of 
the furthest point and the closest point Voronoi diagrams. However, 
in many cases the naive way of making use of this method leads to a 
quadratic algorithm. Consider, for example, the case of the largest
circle separating two simple polygons. Such circle is of one of the two
possible types: it is either tangent to three edges of the external polygon,
or it is tangent to two edges of the external
polygon and one vertex of the internal one. The circle of the first type may
be found in $O(n \log m)$ time, considering Voronoi circles centered at vertices
of $Vor(Q)$, the closest point Voronoi diagram of the external polygon,
and localizing their centers in $FSVor(P)$, the
furthest site Voronoi diagram of the internal polygon. To find the 
separating circle of the second type, we may superimpose $Vor(Q)$ and
$FSVor(P)$.  Taking into consideration, one by one, each portion of an edge of 
$Vor(Q)$, lying in some face of $FSVor(P)$, leads to the investigation of all
the candidate circles of the second type. However, such structure needs
$O(mn)$ space. We strongly believe, that the largest circle separating 
two simple polygons may be found in better than quadratic time.

Using a convex distance to compute the Voronoi diagram,
our method can be adapted to answer separation queries for
separating curves which are the homothets of a given convex curve.

It is natural to try to extend our approach to higher dimensions. The method
from \cite{okm-ccs-86}, detecting the spherical separability of two sets of
points, is based on linear programming and it gives $O(n)$ solution in any
dimension. However, the paraboloid
transformation method, used in \cite{okm-ccs-86},
seems not applicable in the case of simple polygons. Our algorithm achieves
the linear bound scanning two structures: (1) the list of edges of one polygon,
and (2) a path in the furthest point Voronoi diagram of the vertices of
another polygon. The solution circle is always centered on this path. In the
three-dimensional space, the center of the separating sphere may not belong
to a Voronoi edge of either of the two polyhedra. Our "edge-marching" approach
is then not directly applicable to the higher dimensional case.

It is also tempting to ask for the solutions of the higher-dimensional
version of the problems from section 4. The single point queries may be solved
in $O(\log n)$ time by the similar, point location approach. The separating
spheres are tangent to two or three polyhedral faces. The cells are separated
by parts of disks, orthogonal to polyhedral edges, as well as spherical and
conical surfaces. However, it is not clear how to answer queries involving
two or more points.

\paragraph{Acknowledgments.} Authors thanks the anonumous referees
for they helpfull comments which improve the clarity of this paper.

\newcommand{\etalchar}[1]{$^{#1}$}


\begin{thebibliography}{ABO{\etalchar{+}}89}

\bibitem[ABO{\etalchar{+}}89]{abosy-fmcnp-89}
A.~Aggarwal, H.~Booth, J.~O'Rourke, Subhash Suri, and C.~K. Yap.
\newblock Finding minimal convex nested polygons.
\newblock {\em Inform. Comput.}, 83(1):98--110, October 1989.

\bibitem[AGSS89]{agss-ltacv-89}
A.~Aggarwal, Leonidas~J. Guibas, J.~Saxe, and P.~W. Shor.
\newblock A linear-time algorithm for computing the {Voronoi} diagram of a
  convex polygon.
\newblock {\em Discrete Comput. Geom.}, 4(6):591--604, 1989.

\bibitem[BCD{\etalchar{+}}94]{bcdry-ctbc-94}
Jean-Daniel Boissonnat, Jurek Czyzowicz, Olivier Devillers, Jean-Marc Robert,
  and Mariette Yvinec.
\newblock Convex tours of bounded curvature.
\newblock In {\em Proc. 2nd Annu. European Sympos. Algorithms}, volume 855 of
  {\em Lecture Notes Comput. Sci.}, pages 254--265. Springer-Verlag, 1994.
\newblock to appear in CGTA.

\bibitem[BG95]{bg-aoscf-95}
H.~Br{\"o}nnimann and M.~T. Goodrich.
\newblock Almost optimal set covers in finite {VC}-dimension.
\newblock {\em Discrete Comput. Geom.}, 14:263--279, 1995.

\bibitem[Bha88]{b-cspps-88}
B.~K. Bhattacharya.
\newblock Circular separability of planar point sets.
\newblock In G.~T. Toussaint, editor, {\em Computational Morphology}, pages
  25--39. North-Holland, Amsterdam, Netherlands, 1988.

\bibitem[DJ90]{dj-cmcnp-90}
G.~Das and D.~Joseph.
\newblock The complexity of minimum convex nested polyhedra.
\newblock In {\em Proc. 2nd Canad. Conf. Comput. Geom.}, pages 296--301, 1990.

\bibitem[DK85]{dk-ladsc-85}
D.~P. Dobkin and D.~G. Kirkpatrick.
\newblock A linear algorithm for determining the separation of convex
  polyhedra.
\newblock {\em J. Algorithms}, 6:381--392, 1985.

\bibitem[EP88]{ep-mps-88}
H.~Edelsbrunner and F.~P. Preparata.
\newblock Minimum polygonal separation.
\newblock {\em Inform. Comput.}, 77:218--232, 1988.

\bibitem[Fis86]{f-spcrd-86}
S.~Fisk.
\newblock Separating points by circles and the recognition of digital discs.
\newblock {\em IEEE Trans. Pattern Anal. Mach. Intell.}, 8(4):554--556, 1986.

\bibitem[KA84]{ka-dd-84}
C.~E. Kim and T.~Anderson.
\newblock Digital disks.
\newblock {\em IEEE Trans. Pattern Anal. Mach. Intell.}, PAMI-6(5):639--645,
  1984.

\bibitem[Kir83]{k-osps-83}
D.~G. Kirkpatrick.
\newblock Optimal search in planar subdivisions.
\newblock {\em SIAM J. Comput.}, 12(1):28--35, 1983.

\bibitem[Lay71]{l-sss-71}
S.R. Lay.
\newblock On separation by spherical surfaces.
\newblock {\em Amer. Math. Monthly}, 78:1112--1113, 1971.

\bibitem[Lee83]{l-fchsp-83}
D.~T. Lee.
\newblock On finding the convex hull of a simple polygon.
\newblock {\em Internat. J. Comput. Inform. Sci.}, 12:87--98, 1983.

\bibitem[Meg84]{m-lpltw-84}
N.~Megiddo.
\newblock Linear programming in linear time when the dimension is fixed.
\newblock {\em J. ACM}, 31:114--127, 1984.

\bibitem[Mou92]{m-idssp-92}
D.~M. Mount.
\newblock Intersection detection and separators for simple polygons.
\newblock In {\em Proc. 8th Annu. ACM Sympos. Comput. Geom.}, pages 303--311,
  1992.

\bibitem[MS95]{ms-sapo-95}
Joseph S.~B. Mitchell and Subhash Suri.
\newblock Separation and approximation of polyhedral objects.
\newblock {\em Comput. Geom. Theory Appl.}, 5:95--114, 1995.

\bibitem[OKM86]{okm-ccs-86}
J.~O'Rourke, S.~R. Kosaraju, and N.~Megiddo.
\newblock Computing circular separability.
\newblock {\em Discrete Comput. Geom.}, 1:105--113, 1986.

\end{thebibliography}
\end{document}